# BSK-WBSN: BIOMETRIC SYMMETRIC KEYS TO SECURE WIRELESS BODY SENSORS NETWORKS

Samira Mesmoudi and Mohammed Feham

STIC Laboratory, University of Tlemcen, Algeria
`{s_mesmoudi, m_feham}@mail.univ-tlemcen.dz`

## ABSTRACT

*The Wireless Sensors Network (WSN) is an emergent technology resulting from progress of various fields. Many applications of networks WSN are born. One of the applications which have an operational effectiveness relates to the field of health and allows a medical remote support. Miniature wireless sensors, strategically placed on the human body, create a Wireless Body Sensor Network (WBSN) which allows supervising various essential biological signals (rate of heartbeat, pressure, etc). The sensitivity of medical information requires mechanisms of safety. This performance constitutes a challenge for WBSN because of their limitation in resources energy and data-processing. In this paper we propose a new approach to symmetric cryptographic key establishment, based on biometrics physiology. This approach takes into account WBSN constraints and its topology.*

## 1. INTRODUCTION

Great importance has been given recently to the information technology and its use for improving health services. Significant advances in wireless sensor networks with parallel advances in sensors, made a very important impact on telemedicine.

The appropriate proper integration of medical sensors in health care systems allows doctors to diagnose, monitor and treat patients remotely. In this context the wireless body sensors network (WBSN) has been proposed recently.
In any information system, it is essential to establish a security mechanism to protect information because it is susceptible to fraudulent attacks one or the other when it is stored or when it is transmitted, such as the level of security depends on the particular application.

The elderly population and the increasing number of people with chronic diseases already weigh heavily on health systems worldwide. Fortunately, advances in the field of wireless communications now allow continuous monitoring of individuals in real time, which in many cases, contributes to clinical improvement and to remote control of the users to reduce costs and wasted time for patients, nursing staffs and health centers. Wireless Body Sensors Networks WBSN are already the object of a known acronym that includes scenarios in which several sensors and actuators are located at or near the human body to measure such different physiological parameters in different parts of the body and raise these measures by radio way (by a wireless personal network that implements ZigBee (802.15.4) or Bluetooth (802.15.1)) to a personal server, which implements a PDA, mobile phone, or a PC at home. This server provides graphical interface or audio to the user, collects and stores received data, and transmits data to a medical server via the Internet network or mobile phone (2G, GPRS, 3G), which allows the doctor to make decision, etc, [1].

Depending on the application scenario, body sensors networks are used in an autonomous or in combination with mobile phones or wireless sensors networks (WSN) (figure 1).

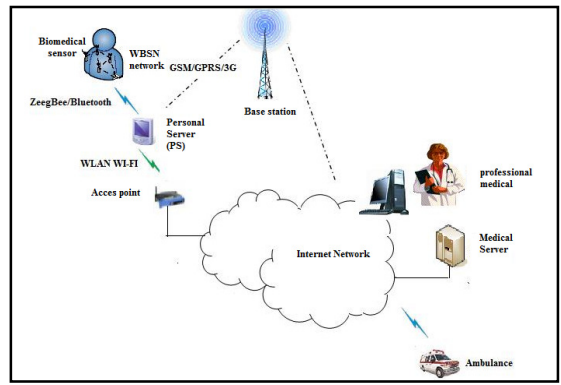

Figure 1: Network architecture of monitoring health system

An autonomous body sensors network consists of small wireless nodes on or near immediate of the patient's body, supplying collectively the function of treatment required by the application. In the simplest scenario, a central node collects and stores the readings of the biosensors such as ECG, EMG, EEG, SpO2 and blood pressure (systolic and diastolic). By providing local treatment measures, the patient is alerted in time when his health changes to the worst.

Topologies based on star and mesh applies to this kind of application class as shown in Figure 2.

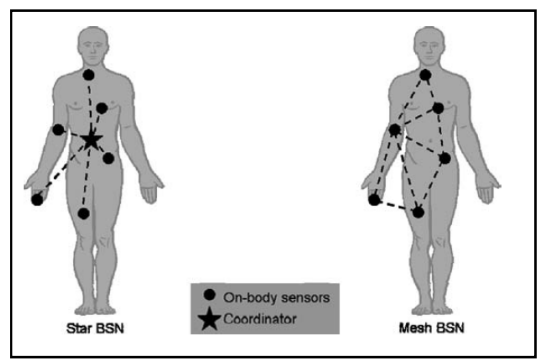

Figure 2: Topology based on star and stitch of the network WBSN [2]

The star topology requires a centralized architecture where the intelligence system is concentrated on a central node which is superior in terms of resources, such as treatment, memory, and energy.

A star network is a common choice. It uses the PDA (Personal Digital Assistant) as local treatment unit to collect and treated the sensor's signals. It is advantageous in situations where a PDA is an inherent part of the system and direct communication between sensors is not required [2].

The concept of distributed system in peer-to-peer networks without central processing unit is used in other projects [2]. Thus the intelligence is moved towards the sensors. Body sensors network consists of intelligent wireless sensors which communicate between them. This is the context of the mesh topology.

Since networks to peer-to-peer does not depend on any particular component, even if a component fails the remaining parts of the system continue to work. This approach is preferable when the sensors must to communicate some with the others.

In this paper we present the WBSN constraints which show the useless of WSN security mechanisms in WBSN. However the WBSN network requires a new solution to take into account these constraints.

In our mechanism, we take in consideration the changeable topology of network to improve the performance of network and increase their level of security. With these conditions we tried to establish three types of keys to secure three types of communications in this network, and we use the biometrics based on physiological data of the human body for the keys distribution.

The rest of this paper is organized as follows. In section 2, we present the related work. The constraints of the wireless body sensors networks and their requirements of the security in section 3. In Section 4, we discuss the security model and how biometric physiological characteristics may increase the level of security in this network. In Section 5, we present our mechanism for establishing symmetric keys, as you take it consideration assumptions on the network in our model solution. Finally, we present the conclusion and future work in section 6.

## 2. RERATED WORK

Relatively a number very limited of works was done in the sector of security of wireless sensors networks. Unfortunately, security mechanisms used in WSN ([3], [4]) can not apply to networks WBSN because BSN nodes must operate with extremely rigorous constraint.

In more the sensors deployed in a WBSN are under the supervision of the person, this means that it is difficult for an attacker to physically access the nodes without it being detected. Therefore network security of WBSN requires appropriate solutions to these problems.

The main security challenges in this network are eavesdropping, injection and modifications of packages and interference of communication between two different networks WBSN.

The medical sensors (nodes BSN) must use cryptographic algorithms to secure the sent data.

Biomedical sensors like any other miniature sensors have constraints on energy and bandwidth.

In more it is also for constraints resulting from their unique location of placement (human body). These constraints play a very important role in the choice of architecture of symmetrical or asymmetrical encryption, key generation method and key distribution protocol is very important for the security level, for example the cryptography by asymmetric key require more resources compared with cryptography by symmetric key in terms of communication and computing.

More relative studies include biometric solution for key distribution ([5], [6]) based on fuzzy commitment [7], specific symmetric cryptographic system with key generation [8], and authentication protocol based on the biometrics [9].

In this work, the physiological signals detected for the health monitoring can be also employed to generate a biometric key ([10], [11]) that can be combined with security mechanisms. Compared to other systems, a higher level of security can be realized with less constraint of calculation and memory. So an algorithm based on biometric data can be used to ensure authenticity, confidentiality and integrity of data transmission between nodes and node to the personal server.

## 3. SECURITY OF NETWORK WBSN

Security in wireless body sensor networks (WBSN) for medical applications is particularly important because sensitive medical information must be protected against unauthorized use that could be dangerous to the life of the user (e.g. change of dosage of drugs or treatment procedures…).

### 3.1. The constraints of security of network WBSN

Some constraints are also experienced by the majority of sensor, but are more rigorous for medical sensors.

Due to the nature of constraints, security solutions proposed for wireless sensors networks do not fit through body sensors networks. Therefore, they need specific solutions. These constraints are:

#### 3.1.1 Low Power

The limitation in power is a major problem for all sensors. But this problem is exceeded in the case of medical sensor. The energy source may be a battery or a rechargeable power source by means of an infrared beam.

The sensors use the power to perform their tasks including the detection of physiological signal, computing and wireless communications. In this process the heat is absorbed by the tissue surrounding the sensor and causing an increase of the temperature. The tissue will also become warm during charging, so the human skin can only tolerate a certain degree of temperature rise that is damaged [5].

#### 3.1.2 Limited Memory

Memory capacity available to medical sensor is very limited due to the size and to the necessary energy consumption. It is of the order of a few kilo-bytes. The implementation of cryptographic protocols does not require much memory, but the storage key, which takes the major part of the memory.

#### 3.1.3 Low computing capabilities

Medical sensors have low computation power, which is limited by two shortcomings: the energy and memory.

Due to the limited memory capacity, it cannot make big calculations; also the most successful function that can be completed by the sensor is the communication of information which was collected. Therefore there is very less quantity of energy that can be consumed in the calculations [5].

### 3.1.4 Low rate of communication

The most expensive operation in terms of energy is the process of communication, so it very important to keep the minimum quantity of communications. It is necessary that these communications which occur for other purposes than data transmission must be minimized.

## 3.2. The security requirements of network WBSN

The data to be transmitted in the current application is medical information. It requires that this information should have the characteristics of secure data authenticity, integrity and confidentiality. The security requirements of body sensors networks are as follows:

### 3.2.1 Confidentiality of data

The data which are communicated between the medical sensors is health information that is personal nature; it should be inaccessible to foreigners (the unauthorized entities). So it is essential that this communication must be confidential.

Confidentiality of data, particularly during transmission when it is vulnerable is realized by encrypting the data by a key.

### 3.2.2 Authenticity of data

Authenticity is the property of the data by which the recipient must be confident that the received signals are resulted from the real sender. This property is very important for body sensors networks because some actions are only launched if the legitimate nodes asked for the action. The absence of this property can lead to situations where an illegitimate entity, disguised as legitimate node, reports data and false instructions to control node (server Personal) possibly causing problems.

### 3.2.3 Data Integrity

The integrity is the property that allows the verification of data. The data must not be altered or modified during transmission because the change of this data can provoke incorrect signals and reactions.

# 4. MODEL OF BIOMETRIC SECURITY

Unlike some well-known biometric characteristics, such as fingerprint, iris, hand geometry, which are models captured on a specific part of the body, the biometric characteristic used in the WBSN is a physiological sign produced by an individual's biological system, for example Heart Rate Variability (HRV).

HRV is a measure of change beat to beat, which is easily available in several types of physiological signals related cardiovascular system, including ECG (electrocardiogram), PCG (phonocardiogram) and PPG (photoplethysmogram).

Up to here the various strategies exploited the biometrics ECG for cryptographic key generation, which is used in diverse mechanisms of network security of WBSN including encryption, authentication protocol, etc...

## 4.1. The key to biometrics

Several notions were given to the biometric key, ASS (Auto Shared Secret) [12], ISS (Intrinsic Shared secret) [13]. In all cases, the biometric key, used in network security protocols, is a binary sequence extracted from physiological data collected by each node WBSN.

Because of characteristics of time variation of physiological data, this key would also depend on time, and it so needs a synchronization request to generate the sensor assembly simultaneously during a specific period to provide a high security level.

A set of biometric keys are generated simultaneously by nodes WBSN during a specific period T, denoted as

$\{x_i \sqsubseteq \{0,1\}^m$ such that $1 \leqslant i \leqslant N\}_T$,

where $m$ is length in bits of the key and $N$ is the network size.

All keys should meet the following requirements ([5], [12]):

Each element $(x_i)$ should possess an unpredictable random characteristic.

Any pair of elements $(x_i, x_j)$ should have high similarity.

The similarity of each pair can be measured by the Hamming distance, and any two nodes will accept each other only if the distance is no more than a threshold $t$ ($t \leqslant m$).

In the following part we distinguish two essential points which are used in a biometric key.

### 4.1.1 Time variation and random

At this point, it is useful to distinguish between time invariant and time-variant biometrics. In most conventional systems, biometrics are time invariant, for example, fingerprints or irises, which do not depend on the time measured. This is based on the recorded biometric, and an authority can identify or authenticate an individual. By contrast, physiological signals (For example ECG) are time-variant.

WBSN security, as most systems, requires variable time biometrics, which makes it a main factor for a high level of security.

Better cryptographic key need an important level of random aspect, and keys derived of signals random (time varying) allow higher security, such as an intruder cannot predict the real key. This is particularly the case with the ECG, since it is time varying, changeable with the diverse physiological activities ([10], [11]).

### 4.1.2 The time of synchronization and recuperation of key

The key aspect of randomness is only part of the security problem. The second condition is that the key generated from physiological signal should be regenerated with high fidelity to the diverse sensor's nodes in the same WBSN.

In particular, the signals captured should be identical independently of the place of measure.

Therefore, to get back identical signals to various sensors, the precise synchronization is a main condition (treated in emission level network).

The system can impose the renewal key so frequently as necessary to satisfy the security demand of the envisaged application: more regeneration the assured security is higher ([10], [11]).

# 5. MECHANISM FOR ESTABLISHING AUTHENTICATED SYMMETRIC KEY

In the previous sections, necessity of assuring the communication in the WBSN and their specific requirements of security were presented. We also explained how all the requirements could be satisfied by means of having a secret key shared between nodes. Once all the communicating entities have the same key, it can be used to execute cryptographic functions such as encryption and MAC (Message Authentication Code).

## 5.1 Hypothesis on the network

In our proposal, we take into account the network topology WBSN, so we use a hybrid topology that combines the two topologies, star and mesh. In this case, we have a leader of the biomedical sensors in the network. This leader has additional operations such as data aggregation and transmission to the base station.
This topology performs decentralized ad hoc operations where no single node is more important than any other in the network.
In our approach, we consider the set of sensors embedded in an individual. This set forms a group with a leader, where sensor nodes are connected with relationships of master-slave type.
In this topology, the nodes broadcast their identifiers and listen to the neighbors; these identifiers are added to its routing table. Every time this group elects a sensor node as its head and all communications inter-group is forwarded by this leader (cluster head).
The cluster head (CH) also serves as a fusion node of aggregates packets, before sending them to the base station [14]. From time to time the CH is changed to increase the lifetime of the network and improve its performance. This cluster head is replaced by another when its energy reaches a minimum value of threshold.
So we tried to establish authenticated symmetric-key and we take into account the change of the leader.

## 5.2 The use of the biometric

The different reading of biometrics are independent some of the others. Hence this situation can be considered as analogous to that in which the error is introduced into the data during transport leading to a non-zero of Hamming distance between the data sent and received. Error correction would help to limit this problem of error. A parameter code $(M, K, D)$, where $M$ is code length, $K$ is the length of the realities biometric data, and $D$ is the minimum distance of code which allows to corrects $e = (D - 1)/2$ errors is appropriate. The number of errors can be reduced by taking multiple readings independently way and using the code obtained by encoding majority of these readings. The fuzzy commitment scheme incorporates error correction codes to protect or encrypt data. There are two phases in this system to find commit phase and decommit phase [5].
In the commit phase of the entity to be protected (for example) $c$ is committed with $x$ as proof using $F_{com}$

$$F_{com}(c, x) = (h(c) \,||\, \delta)$$

Where $\delta = x \oplus c$ ($\oplus$ is the operation XOR bit with bit) and $h()$ is a hash function. The receiver receives $h(c) \,||\, \delta$ (| | is the concatenation operation) of the sender. Now, the receiver decommits $c$ to assistant $F_{dec}\,(h\,(c)\,||\,\delta, x\,')$ as follows. It calculates $c\,' = f\,(x' + \delta)$; is a variant version of $x$ evidence available to the receiver and $f$ is a function of error correction. Now the receiver verifies if $h\,(c) = h\,(c\,')$. If they are equal then the operation is successful.
At this point we can recuperate the entity to protect.

## 5.3 Solution model

In the network WBSN, There are three types of wireless communication links. They are the communication links between biomedical sensors and the leader node, the communication links between biomedical sensors nodes and the link between leader node and the personal server.
In our proposal we are interested to secure the first two types of communications, so we try to establish authenticated symmetric key shared between the leader and other nodes and a key shared by all sensors' nodes.
We propose here to apply biometrics to the key establishment protocol.

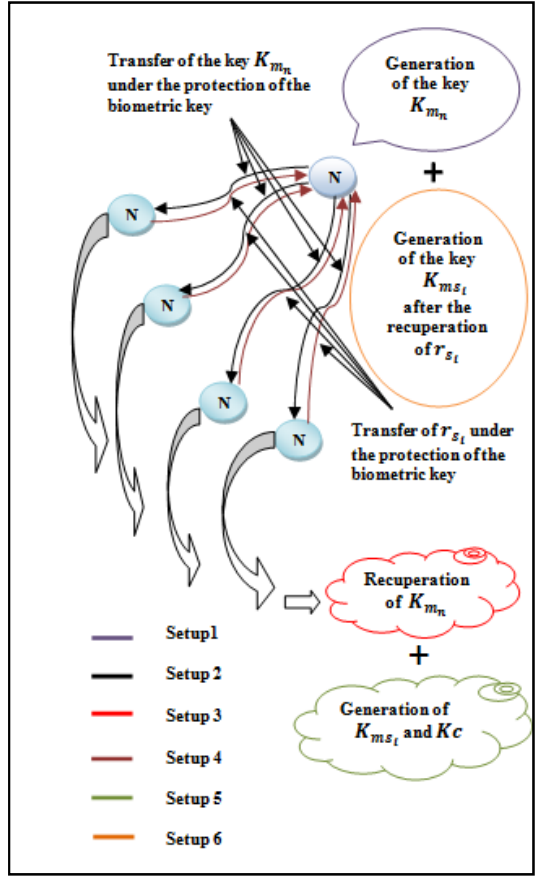

Figure 3: Mechanism of establishment of key

In this protocol (Figure 3), leader (master) initiates the process, first, the key $k_{m_n}$ is generated by the leader $m_n$ size of 128 bits ($m_n$: is the master sensor of index $n$) and transferred to other biomedical sensors $s_i$ ($s_i$: is the sensor slave of index $i$ such that $0 \leq i \leq N$ such that $N$ is the number of sensors in the network) under the protection of the biometric key by using the function Fuzzy Commitment.

$$F_{com}(k_{m_n}, k_{commit}) \parallel IDm_n \parallel Ts$$
$$\parallel MAC(F_{com}(k_{m_n}, k_{commit}) \parallel IDm_n \parallel Ts)$$

Such as $k_{commit}$ is the key which is calculated for each session, it stores a number used to commit the key $k_{m_n}$, it is derived by the combination $x_{m_n}$ and $r_u$

$$k_{commit} \leftarrow x_{m_n} \oplus r_u$$

With

$x_{m_n}$ : the biometric key generated by the leader, its length is 128 bits.
$r_u$: Stores a number which is unique to the individual, i.e. it is shared by the entire sensors in the network. Its length is 128 bits.
In the message sent, $IDm_n$ which represents the identification of leader sensor $m_n$, is unique for each sensor.
$Ts$ : is the timestamp to prove the non-replay of an old message.
The function MAC assures that the message is unchanged (integrity) and comes from the sender (authentication, by using the secret key). It is created from hash functions; it can also be used as an additional encryption. The MAC is calculated on the message sent by using the key $k_{m_n}$ and generates an output of 128 bits

During the reception, the sensor $s_i$ (slave) try to recuperate $k_{m_n}$ using its own $k'_{commit}$, i.e. the use of $x_{S_i}$ (the biometric key generated from the side of biomedical sensor)

$$k'_{commit} \leftarrow x_{S_i} \oplus r_u$$
$$k'_{m_n} = F_{dec}(F_{com}(k_{m_n}, k_{commit}), k'_{commit})$$

If the operation of decommitment is successful, $k'_{m_n} = k_{m_n}$. The receiver node will execute the same calculation of MAC on the message and will compare it with the MAC received, if it does not match, then the packet is rejected.
After recuperation of the key $k_{m_n}$, each biomedical sensor's node generates its appropriate key $k_{ms_i}$.

$$k_{ms_i} \leftarrow k_{m_n} \oplus r_{s_i}$$

Such as $r_{s_i}$ is a unique number that is generated by each sensor, its length is 128 bits. Each generated key is used to secure communications between the leader and a sensor node in the network.
With the same method, each sensor's node sends $r_{s_i}$ to the CH to generate the key $k_{ms_i}$, but here we use $IDs_i$ in the message to send the message, which is the identification of sensor's node sender.

$$F_{com}(r_{s_i}, k_{commit}) \parallel IDs_i \parallel Ts$$
$$\parallel MAC(F_{com}(r_{s_i}, k_{commit}) \parallel IDs_i \parallel Ts)$$

With

$$k_{commit} \leftarrow x_{S_i} \oplus r_u$$

The master node retrieves all $r_{s_i}$ by decommitment operation and controls over the message integrity by verifying the MAC.
So we have two types of keys, the first is $k_{ms_i}$ that is specific for each biomedical sensor's node used to secure communications between the node leader and a biomedical sensor node, and the other is the sensor key $ks$ that is used to secure communications between the sensors themselves.
Such as

$$ks \leftarrow k_{m_n} \oplus r_u$$

What concerns third typical of communications i.e. The communications between the CH and the personal server, to secure communications we use the network key $kn$, which is generated by the personal server, pre-deployed in each sensor nodes.

The personal server uses this key to decrypt the data, and the leader uses this key to encrypt data and send it to the personal server. The biomedical sensor node uses this key only one case where this node is a leader.

**Leader's change of network WBSN**

To maximize the lifetime of network, a node with higher energy is chosen to be a leader because he has to perform several operations such as data aggregation, etc. This rate of energy consumption is very high in case the node is cluster head.
We apply a rotation so that when the energy of a leader node reaches a minimum of threshold, it will transfer responsibility to another node by election.
The steps for selecting the leader in the network are:

Step1: Every time a leader of network estimates that it is incapable to support the responsibility for some reasons, as the low energy, it will broadcast a message for a re-election.
Step2: If a node receives a message of election, it will find a neighbor with the highest energy value and sending to the current leader as a vote.
Step3: Now the leader will assign this responsibility to the node having the largest number of votes.
After the election, how now the new leader takes responsibility and establishes symmetric keys authenticate for the various communications?
Messages received by the leader from the sensor's nodes are forwarded to the new leader. The leader adds his $IDm_n$ for the incoming packet and reconstructs the package to the title. But before transmitting data packets to the new leader, it is necessary to establish again keys for secure communications.
In this case, the former leader sent request messages to the new leader that contains:

$$IDm_n \parallel F_{com}(r_{s_i}, k_{commit}) \parallel IDs_i \parallel Ts$$
$$\parallel MAC(IDm_n \parallel F_{com}(r_{s_i}, k_{commit}) \parallel IDs_i \parallel Ts)$$

Such as
$IDm_n$: is identification of the former leader.
The new leader tries to retrieve all $r_{s_i}$ by the use of decommitment by own sound $k_{commit}$ i.e. to use their biometric key $(x_{m_{n'}})$.
After the recuperation of any $r_{s_i}$ by the new leader, then it generates a new key $k_{m_{n'}}$ on the place of $k_{m_n}$ that is used by the former leader.

$k_{m_{n'}}$ is sent with the same procedure to sensor's nodes and the same steps above are repeated to obtain the keys $ks$ and $k_{ms_i}$. The new leader adds its $(IDm_{n'})$ to send packets to the personal server more than the identification of the former leader for alerting that there is a change. At every change of leader we repeat the same steps above.

## 6. CONCLUSION

In this article we have established the essential characteristics of wireless body sensor networks (WBSN) to show the impossibility to use the security mechanisms implemented in wireless sensor network (WSN) in the network (WBSN) because of their very rigorous constraints compared to (WSN). Nevertheless, the inclusion of security in the development of new solutions had an essential place to guarantee the stability of the system.

We have taken into account the propositions of some authors that are based on the use of physiological data of the human body as a biometric means in the processes of security.

Contrary to the conventional biometric means, the physiological data have specific properties as changing with time and high level of randomness. This proposal can lead to reasonable solutions and accessible to various constraints of the WBSN.

So, we have presented a mechanism for establishing cryptographic authenticated symmetric key, we rely on the biometric characteristic of the physiological signal, and we take into account the network topology and the change in the real time. Our aim is to improve the security level and network performance.

The future work involves examining the performance of the key biometrics so that their use gives a very important impact for the level of security. The implementation intervenes finally to test the efficiency of this mechanism in the wireless body sensor networks.

**Authors**

**Samira Mesmoudi,** received her engineer degrees in Telecommunication from the University of Tlemcen, Algeria in 2006, and her M.S. degrees in networks and telecommunication systems from the same University. She is a PhD candidate at the University of Tlemcen. Her research interests include wireless sensor network, wireless body sensor network and their security.

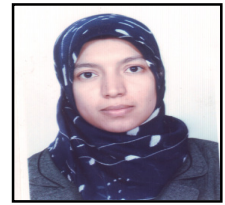

**Mohammed Feham,** received his PhD in optical and microwave communications from the University of Limoges, France in 1987, and his PhD in science from the University of Tlemcen, Algeria in 1996. Since 1987 he has been assistant professor and professor of microwave and wireless communication. His research interest is mobile networks and services.

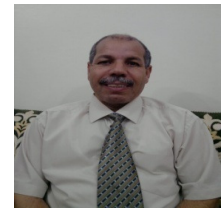